\documentclass[12pt,preprint]{aastex}
\usepackage{emulateapj5,onecolfloat,epsfig}

\def\vi{$V_{606}-I_{814}$\/ }
\def\vir{$V_{606}-I_{814}(r)$\/ }
\def\dvi{$\delta(V-I)$\/ }

\slugcomment{submitted to ApjL May 23, 2001}

\shorttitle{The origin of Blue Cores in HDFs spheroidals}
\shortauthors{F. Menanteau, R. Jimenez \& F. Matteucci}

\begin{document}

\twocolumn[ 

\title{The Origin of Blue Cores in Hubble Deep Fields E/S0 galaxies.}

\author{Felipe Menanteau\altaffilmark{1}} 
\affil{Carnegie Observatories, 813 Santa Barbara Street, Pasadena, CA 91101}
\email{felipe@ociw.edu}

\author{Raul Jimenez}
\affil{Dept of Physics \& Astronomy, Rutgers University, 136
Frelinghuysen Road, Piscataway NJ 08854}
\email{raulj@physics.rutgers.edu}

\author{Francesca Matteucci}
\affil{Dipartamento di Astronomia, Universita di Trieste, via G.B. Tiepolo 11,
  34131 Trieste, Italy}
\email{francesc@ap.sissa.it}

\begin{abstract}
  In this letter we address the problem of the origin of blue cores
  and inverse color gradients in early-type galaxies reported in the
  Hubble Deep Field North and South (HDFs) by
  \cite*{Menanteau-etal-01}. We use a multi-zone single collapse
  model.  This model accounts for the observed blue cores by adopting
  a broad spread in formation redshifts for ellipticals, allowing some
  of these galaxies begin forming no more than $\sim1$ Gyr before the
  redshift of observation. The single-zone collapse model then
  produces cores that are bluer than the outer regions because of the
  increase of the local potential well toward the center which makes
  star-formation more extended in the central region of the galaxy. We
  compare the predicted \vir color gradients with the observed ones
  using the redshift of formation ($z_F$) of the elliptical as the
  only free parameter. We find that the model can account with
  relatively good agreement for the blue cores and inverse color
  gradients found in many spheroidals and at the same time for the red
  and smooth colors profiles reported. Based on the model our analysis
  suggests two populations of field ellipticals, one formed recently,
  within $\lesssim1$Gyr and another much older formed $\gtrsim4$Gyr
  since the redshift of observation.

\end{abstract}


] 

\altaffiltext{1}{Gemini Fellow}

\section{Introduction}

The study of field elliptical galaxies has seen a dramatic boost in
the past couple of years, mainly from the realization that these
systems can be objectively used to test opposed models of galaxy
formation.

For a number of years it was accepted the hypothesis that field
elliptical galaxies formed in isolation during a single burst of
formation at high redshift.  This had its origin in the early
monolithic collapse model postulated by \cite{Eggen-etal-62} and later
sustained by observations of ellipticals in rich clusters over a wide
redshift range which closely follow a number of ``fundamental
relations''. The low scatter in the fundamental plane
\citep{Djorgovski-Davis-87, Dressler-etal-87} and in the
color-magnitude relation (CMR) \citep{Sandage-etal-78, BLE92,
Ellis-etal-97, vanDokkum-etal-98} implied a high degree of homogeneity
in the stellar population, both consistent with a model in which
ellipticals formed at high redshift and since then passively evolved.

On the other hand, in hierarchical CDM models \citep{White-Rees-78,
White-Frenk-91} galaxies are formed only as the result of merging of
smaller sub-units, suggesting that massive ellipticals could only have
been assembled recently (since $z\simeq1$) and in consequence there
would be a paucity of early-types at higher redshifts \citep{KCW96}.
Unfortunately, much of the observations were early focused on
elliptical galaxies in rich clusters, where the existence of old
ellipticals at high redshift is expected in hierarchical CDM models
because evolution is accelerated in these dense
regions \citep{Governato-etal-98,Kauffmann-etal-99b}.

\begin{figure}[t!]
  \centerline{\epsfig{file=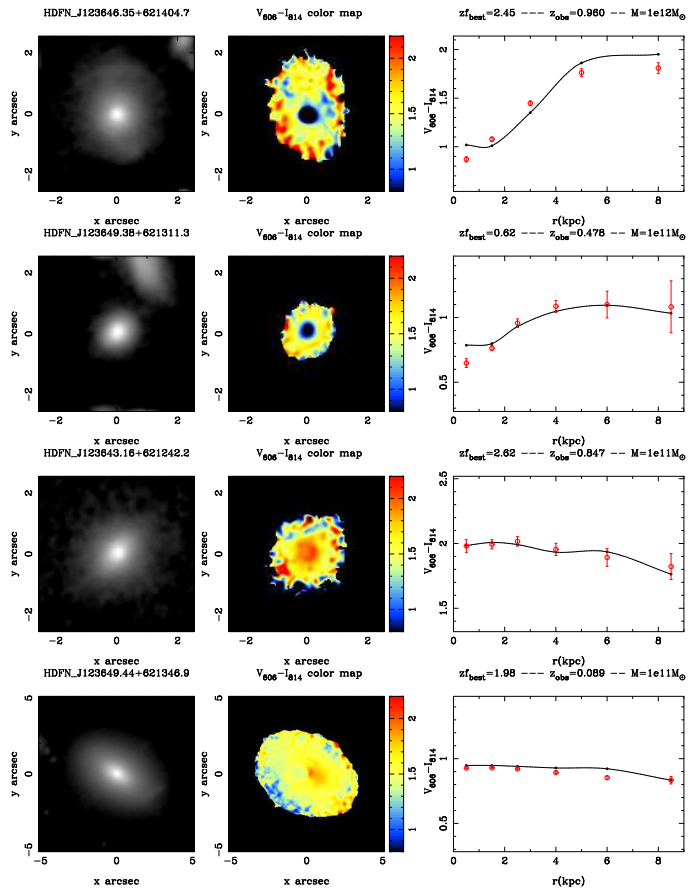,width=9cm}} 
  \caption{An example of the ability of the model to reproduce a range
  of different color gradients. Figure shows from left to right:
  $I_{814}$--band surface brightness map, \vi color pixel map and \vir
  color gradient for several E/S0 in the sample. Open circles
  represent observed gradients while solid lines the model
  predictions. The complete set of fits can be found at {\tt
  http://www.ociw.edu/$\sim$felipe/blue-cores}}
   \label{fig:example}
\end{figure}

Several studies have attempted to test whether field ellipticals were
formed at high redshift and evolved in isolation since, following to
some extent the methodology from \cite{Moustakas-etal-97} and
\cite{Zepf-97}, consisting of using a combination of optical and
infrared colors for selecting spheroidals in search for a decline in
the co-moving density of spheroidals; all of them with consistent
results, a significant lack of red objects compared with the
predictions of pure luminosity evolution (PLE) models
\citep{Barger-etal-99, Daddi-etal-00}. The addition of HST imaging
lead to the conclusion that a model with constant co-moving density
and a single epoch of formation for E/S0 at {\em all} epochs is ruled
out. A more conclusive test in this line will need a complete census
of ellipticals up to $z=1$, recent results from \cite{Im-etal-00} are
inconclusive but compatible with very mild density
evolution. Similarly, using dataset multicolor information it has been
reported a population of E/S0 with colors significantly bluer than
passive evolution \citep{KBB99,Abraham-etal-99}.
In a recent detailed analysis \cite*{Menanteau-etal-01} reported,
using resolved multicolor photometry of E/S0 in both HDF-North and
South, solid evidence for the continuous star-formation in E/S0
galaxies. The resolved $(V-I)$ data revealed a population of
spheroidals with central blue cores, color inhomogeneities and inverse
color gradients, which were indicative of formation activity at the
galaxies redshift of observation.  Interestingly, a subset of
spheroidals had the color properties of old stellar population in
good agreement with the ones prescribed with a stellar population
formed in a single burst at $z>2$.

In recent years, the hierarchical CDM semi--analytical models have
dominated the predictions in the field leaving alternative frameworks
relatively unexplored in which to interpret data. The realization that
spheroidals, even at intermediate redshifts, are a mixture of both a
mild star-forming and old stellar population, makes imperative to
search for alternative formation models which can successfully
reproduce both properties, the blue cores and smooth color profiles
observed. In the past, some authors have investigated the effect of
adding small amounts of SF activity to account for some of the blue
colors observed in E/S0 (e.g. \citealt{Menanteau-etal-99},
\citealt{Trager-etal-2000}). Although they have been successful to
some extent in reproducing the observed colors, they rely on somewhat
ad-hoc recipes.

Recently, \cite{Friaca-Terlevich-01} have used a multi--zone
chemo--dynamical model to predict the color gradients of spheroids at
different redshifts. In their model an important component to
reproduce the blue cores is a late infall of gas that extents the
duration of the star formation. In this letter we investigate a
similar model but without the late infall feature and take it a step
further by producing detailed fits to the observed color gradients of
all 77 E/S0 studied by \cite*{Menanteau-etal-01} in order to asses
whether such single--collapse models really fit the observations at
all radii and not only the qualitative trend of a blue core. We show
that the color gradients are determined by the shape of the
gravitational potential and therefore the different escape velocities
of the gas rather than the need of late infall. In our model the
majority of the gas ($>80\%$) is assembled at a very early epoch in
the formation of the elliptical. This model could be representative of
how ellipticals assemble in a warm dark matter scenario (where small
scale structure is suppressed) or in a CDM scenario in which the
subsequent merging of small objects on a big galaxy has been
suppressed --maybe due to the lack of small scale power (e.g.
self--interacting dark matter) or strong feedback from the big central
galaxy.

\section{The HDF North and South E/S0 Sample}

The E/S0 used in our analysis are taken from the field sample of
galaxies selected by \cite*{Menanteau-etal-01} from the HDFs. Here we
include a brief description of it, but we point the reader to
\cite*{Menanteau-etal-01} for a complete description of it. The sample
consists of 77 spheroidal galaxies, classified as E, E/S0 and S0 under
the Hubble scheme with a limit near-total magnitude of
$I_{814}=24$~mag. Throughout this {\em letter} magnitudes are in the
Vega system. Source detection and integrated photometry were taken
from Sextractor catalogs constructed by the Space Telescope Science
Institute following the IAU IDs given there. The morphological
selection of E/S0 was performed using both visual and machine based
classification in order to ensure a robust morphology selection (see
\citealt*{Menanteau-etal-01}). Redshift information was taken from
publicly available compilations, and augmented with photometric
estimates provided by S. Gwyn (private communication).

\subsection{Observed color gradients}

The quantity that we use to compare the properties of spheroidals with
the model predictions (see next section) is the color gradient of
galaxies. We focus only on the $V_{606}-I_{814}$ colors to estimate
the color gradient $V_{606}-I_{814}(r)$ as a function of the radius
$r$, given that these represent the bands with higher signal-to-noise
in both HDFs. In order to obtain the color gradients, we first compute
the centroid and second order moments ($\overline{x^2}$,
$\overline{y^2}$, $\overline{xy}$) of the galaxy utilizing the
$I_{814}$--band, from which we estimate the ellipticity parameters of
the objects ($A, B, \theta$). It is worth noting that we keep the same
elongation ratio $e = A/B$ up to the isophotal limit to which the
color gradient is computed. Finally using concentric ellipses, we
compute the $V_{606}-I_{814}(r_i)$ gradients as the median color in
the shell between $r_{i-1}$ and $r_{i+1}$.

\section{Making an Elliptical}

We model elliptical galaxies as a system with spherical symmetry and
multiple zones. In particular it is assumed that the bulk ($> 80\%$)
of the gas in this model was in place at the time of formation and was
able to form stars, i.e.  was cool enough. The galaxy is then divided
in spherical shells --100 for the present case-- each of them
independent, i.e. no transfer of gas is allowed among shells. In each
of these shells star formation proceeds according to a Schmidt law:
SFR$=\nu \rho_{\rm gas}(t)$, where $\rho_{\rm gas}(t)$ is the volume
gas density in the shell and $\nu=8.6(M_{\rm
gas}/10^{12}M_{\odot})^{-0.115}$Gyr$^{-1}$. The initial mass function
is assumed to be a power law ($\phi(m) \propto m^{-0.95}$). Star
formation proceeds in each shell until the gas is heated up by SN to a
temperature $T$ that corresponds to the escape velocity of {\it each}
shell (see \citealt*{Matinelli-Matteucci-Colafrancesco-01} for a
detailed description of the model). The gas in the elliptical galaxy
is assumed to be within a dark matter halo of mass 7 times larger than
the gas mass ($\Omega_m=0.35$ and $\Omega_b=0.05$). The dark matter
follows the density profile described in
\cite*{Matinelli-Matteucci-Colafrancesco-01}. The chemical enrichment
of the gas and stars is followed in detailed using the stellar yields
of \cite{Woosley-87} for massive stars and supernovae II
($M>8M_{\odot}$), \cite{Renzini-Voli-81} for low and intermediate mass
stars ($0.8 \le M/M_{\odot} \le 8$) and
\cite*{Thielemann-Nomoto-Hashimoto-93} for SN type Ia; we also took
into account the lifetimes of stars. For each shell we assume that
mixing of the gas is very efficient in the whole shell and shorter
than the lifetime of the most massive stars.

For different masses the model predicts a different time dependence of
the SFR. In more massive systems the potential well will be deeper and
therefore it will take longer for the gas to reach temperatures larger
than the escape velocity in the potential, thus star formation will
last longer than in less massive systems. Also, for a fixed mass,
since the potential is deeper in the core of the galaxy, the model
predicts that star formation will last longer in the core than in the
outer regions (see \citealt*{Matinelli-Matteucci-Colafrancesco-01}).
The model provides us with the mass fraction that is turned into stars
at a given time and their chemical composition. This output is then
fed into a synthetic stellar population code \citep{Jimenez-etal-98}
to compute the spectral properties of the galaxy at different radii
and times.

\section{Results}

\begin{figure}[t!]
  \centerline{\epsfig{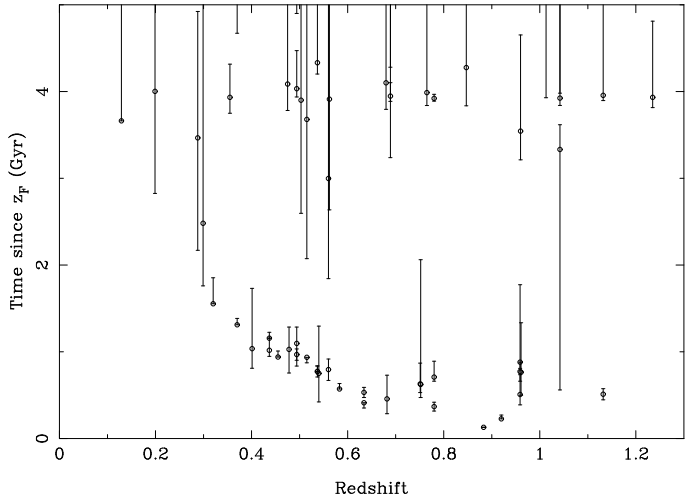}}
  \caption{Time since the most likely redshift of formation ($z_F$)
  for all E/S0 in the HDFs sample as a function of its redshift of
  observation. $z_F$ is calculated as described in the text. Error
  bars represent 1$\sigma$ values on the $\chi^2$
  fitting.}
  \label{fig:time-zf}
\end{figure}

The aim of our analysis is to directly compare the observed \vir color
gradients with the predictions of the above model. In the first place
we are interested to see whether the color gradients can be reproduced
at all. We note that the only free parameter now in our model is the
redshift of formation given that mass is determined by the brightness
of the galaxy at the redshift of observation. To accomplish this we
use a maximum likelihood method ($\chi^2$) to compute the most-likely
redshift of formation ($z_{F}^{best}$) for the best-fitting
model. In order to avoid spurious results from small fluctuations in
the gradients due to noise, the model and data gradients were
re-binned to a common grid of $5-6$ shells up to a maximum physical
radius of $10$~kpc. In order to transform the observed color gradients
to physical (kpc) length, we assumed a flat cosmology with
$\Omega_m=0.35$ and $H_0=65$ km s$^{-1}$ Mpc$^{-1}$. Finally, as the
HST point spread function (PSF) is wavelength dependent, in order to
properly compare observed and modeled gradients, we convolved each
modeled surface brightness profile with a 1D radial Gaussian PSF of
different FWHM in the $V_{606}$ and $I_{814}$--band
\citep{Casertano-etal-00}.

We applied this methodology to the whole sample of 77 E/S0
galaxies. To illustrate the method, in Figure~\ref{fig:example} we
show a selection of four representative galaxies in the sample. We can
see how the model can successfully account for the observed range of
\vir color gradients. The complete set of fits can be found at {\tt
http://www.ociw.edu/$\sim$felipe/blue-cores}. It is worth noting the
ability of the model to reproduce the blue cores (inverse steep
gradients) present in the sample (Figure~\ref{fig:example} upper two
panels), both the color difference and the physical scale at which
this occurs. In addition very flat and smooth color profiles can be
successfully reproduced (Figure~\ref{fig:example} lower two panels) as
expected from a stellar population formed in single burst at high
redshift. To emphasize the power of our model, we focus on the bottom
panel elliptical at $z=0.089$. This exhibits a slightly red color
profile induced by a metallicity gradient, which the model naturally
accounts for due to the different rate at which star formation
proceeds in each shell yielding to a metallicity gradient.

The redshift of formation is the relevant measurement in our
analysis. This is of importance to understand when E/S0 assembled to
their final morphological shape in which we see them
today. Figure~\ref{fig:time-zf} shows the time since $z_{F}^{best}$
(computed galaxy redshift of formation) as a function of the observed
redshift for the whole sample. The most striking feature is the
bimodal distribution of times: a population seems to be formed within
$1$~Gyr, while a second population seems to have formed at least
$4$~Gyr since they were observed. We note that for the later
population the time since formation could in principle be bigger than
$4$~Gyr, since the colors at old ages do not contain enough
information as to give a precise measurement of the redshift of
formation. This population, which the model predicts was assembled at
an early epoch, agrees with the passive evolving population detected
independently under the basis of \dvi color scatter by
\cite*{Menanteau-etal-01}. It is worth to notice that the model yields
a relatively low redshift of formation for some galaxies when they
have large inverse color gradients or marked blue colors in order to
account for them.

\begin{figure}[t]
  \centerline{\epsfig{file=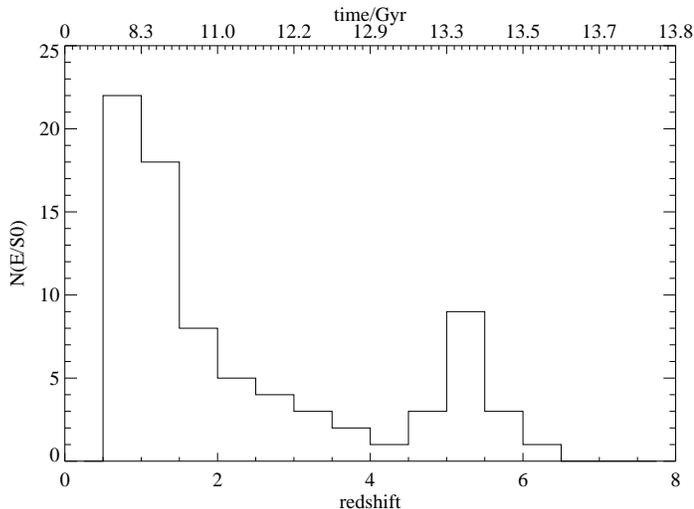,width=9cm}}
  \caption{Histogram for the most likely redshift of formation, $z_F$
   values for the whole sample of E/S0.\label{fig:hist}}
\end{figure}

Figure~\ref{fig:hist} shows the distribution of the formation redshift
for the whole sample. The top axis shows the look--back--time. About
25\% of field E/S0 in our sample have formed at $z\gtrsim4$, with
$\sim 30\%$ of the sample having formed at $z<1$. The medium redshift
of formation is $z \sim 2$ and therefore the medium age of the field
ellipticals in the HDFs is $\sim 11$ Gyr. Therefore, as a whole, field
ellipticals are only 1--2 Gyr younger than cluster ellipticals, in
agreement with the findings by \cite{Bernardi-et-al-98} who compared
the Mg$_2-\sigma_0$ relation for field and cluster ellipticals. The
main feature of Fig.~\ref{fig:hist} is the continuous formation of
field E/S0 with redshift, we do not find evidence for a {\it single}
epoch of spheroid formation. Although for the reasons discussed above
the redshift of formation can be quite uncertain at the high--redshift
end, it is clear that we can distinguish a trend of a bimodal
distribution in the formation redshift. This may be indicative of two
preferred epochs of formation, according to the model. However,
this result should not be over interpreted, as this is a model
dependent one based on the success of the model to reproduce the
observations. 

\section{Conclusions}

In this letter we have presented an alternative approach for the
formation of E/S0 galaxies. Recognizing the features that E/S0 show at
low and intermediate redshift our analysis seeks to exploit the
resolved information now available for E/S0 to quite high redshift
focusing on their color gradients as the basis of our analysis. In
order to obtain information from the color--gradients about the
formation epoch of E/S0, we have adopted a multizone single collapse
model for the formation of spheroids. In the context of this model,
our main conclusion is that E/S0 can be properly described by a model
in which 80\% of the gas was processed at an early stage, but in which
spheroidals are formed continuously as a function of
redshift. Although there is no preferred epoch for their formation,
our analysis suggest the presence of two population, one assembled
recently, within $\sim1$Gyr and another with the properties of an old
stellar population formed $\gtrsim4$Gyr from the redshift of
observation.  It remains to be learned what star formation histories
can be obtained in a model independent fashion from the spectra of
these objects and whether this yields to similar conclusions.


\end{document}